%% file: timeslicing-short.tex
\documentclass[preprint]{vgtc}               

\ifpdf
  \pdfoutput=1\relax                   
  \usepackage{graphicx}                
  \DeclareGraphicsExtensions{.pdf,.png,.jpg,.jpeg} 
\else
  \usepackage{graphicx}                
  \DeclareGraphicsExtensions{.eps}     
\fi%

\graphicspath{{figures/}{pictures/}{images/}{./}} 

\usepackage{microtype}                 
\PassOptionsToPackage{warn}{textcomp}  
\usepackage{textcomp}                  
\usepackage{mathptmx}                  
\usepackage{times}                     
\usepackage{cite}                      
\usepackage{tabu}                      
\usepackage{booktabs}                  

\usepackage{microtype}                 
\PassOptionsToPackage{warn}{textcomp}  
\usepackage{textcomp}                  
\usepackage{mathptmx}                  
\usepackage{times}                     
\usepackage{tabu}                      
\usepackage{booktabs}                  
\usepackage{algorithm}
\usepackage{algorithmicx}
\usepackage{algpseudocode}

\usepackage{footnote}
\usepackage{url}
\usepackage{graphicx}
\usepackage{paralist}
\usepackage{lipsum}
\usepackage{enumitem}
\setlist[enumerate]{itemsep=0mm}

\usepackage{amsmath}
\usepackage{nccmath}
\usepackage{color}
\usepackage{multirow}
\newcommand{\yongwang}[1]{\textcolor{black}{#1}}
\newcommand{\yongwangblue}[1]{\textcolor{black}{#1}}

\onlineid{1035}

\vgtccategory{Technique/Algorithm}

\vgtcinsertpkg

\preprinttext{To appear in the proceedings of IEEE VIS 19 Short Paper}

\title{Nonuniform Timeslicing of Dynamic Graphs Based on Visual Complexity}

\author{Yong Wang\thanks{e-mail: ywangct@cse.ust.hk}\\ %
        \scriptsize HKUST %
\and Daniel Archambault \thanks{e-mail: d.w.archambault@swansea.ac.uk}\\ %
     \scriptsize Swansea University %
\and Hammad Haleem \thanks{e-mail: hammadhaleem@gmail.com}\\ %
\scriptsize HKUST %
\and Torsten Moeller \thanks{e-mail: Torsten.moeller@univie.ac.at}\\ %
	\scriptsize University of Vienna %
\and Yanhong Wu \thanks{e-mail: wdan1990@gmail.com }\\ %
	\scriptsize Visa Research %
\and Huamin Qu \thanks{e-mail: huamin@cse.ust.hk}\\ %
	\scriptsize HKUST} %

\teaser{
	\centering
	\includegraphics[width=0.92\linewidth]{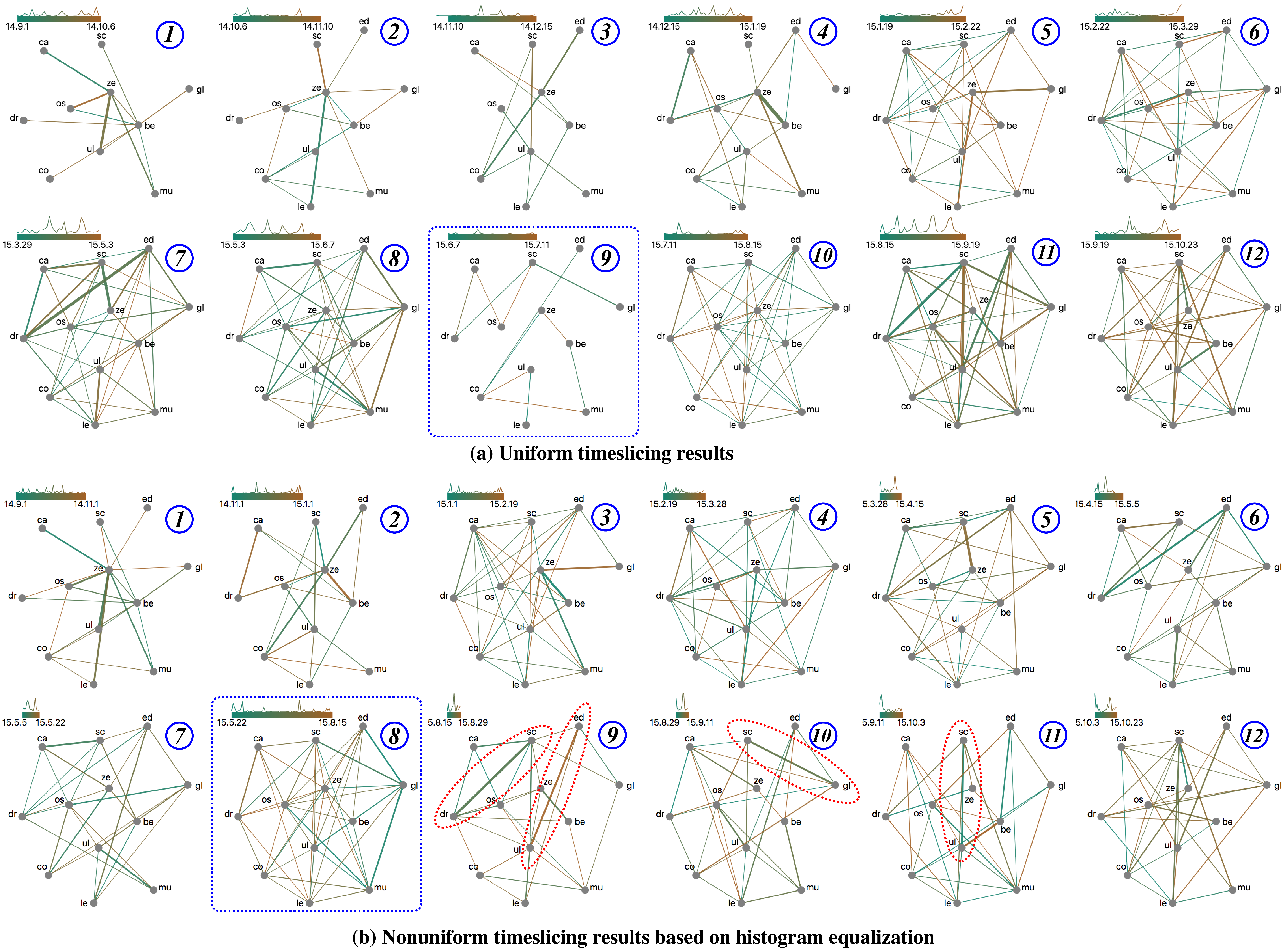} 
	\vspace{-1em}
	\caption{Comparison of uniform timeslicing and non-uniform timeslicing using the Rugby Dataset. (a) Traditional uniform timeslicing, (b) non-uniform timeslicing. The whole stream of temporal edges are divided into 12 intervals with the sequence number marked at the top right corner of each snapshot. The top left bars and smoothed line chart show the time range of each interval (in a format of ``\emph{year}.\emph{month}.\emph{day}'') and the edge frequency distribution, respectively. The dotted blue rectangles highlight two interesting intervals and the red dotted ellipses highlight several thick edges. The teams are: \textit{ca - Cardiff, sc - Scarlets, dr - Dragons, os - Ospreys, le - Leinster, mu - Munster, ul - Ulster, co - Connacht, ed - Edinburgh, gl - Glasgow, ze - Zebre, and be - Benetton}. }
	\label{fig-rugby-case-study}
	\vspace{-0.4em}	
}

\abstract{Uniform timeslicing of dynamic graphs has been used due to its convenience and uniformity across the time dimension. However, uniform timeslicing does not take the data set into account, which can generate cluttered timeslices with edge bursts and empty timeslices with few interactions. 
The graph mining filed
has explored nonuniform timeslicing methods specifically designed to preserve graph features for mining tasks. 
\yongwang{
In this paper, we propose a nonuniform timeslicing approach for dynamic graph visualization. Our goal is to create timeslices of equal visual complexity. To this end, we adapt histogram equalization to create timeslices with a similar number of events,
balancing the visual complexity across timeslices and conveying more important details of timeslices with bursting edges.
A case study has been conducted, in comparison with uniform timeslicing, to demonstrate the effectiveness of our approach. }
} 

\CCScatlist{
  \CCScatTwelve{Human-centered computing}{Visu\-al\-iza\-tion}{Visu\-al\-iza\-tion techniques}{Graph Visualization};
}

\begin{document}

%
%
%
%

\input{intro}

\input{related_work}

\input{prob_def}
\input{method}
\input{case_study}

\input{conclusion}


\acknowledgments{
\yongwangblue{This work is partially supported by grant RGC GRF 16241916.}
}

\bibliographystyle{abbrv-doi}

\bibliography{reference}
\end{document}

%% file: intro.tex
	
\firstsection{Introduction}
\label{sec:introduction}

\maketitle
	

%
%
%
%
%
%
%
%
%
Graphs are widely used to represent the relations between different objects.
\yongwang{
Many of these graphs dynamically change over time and are ubiquitous across various applications and disciplines, such as social networks, (tele-)communication networks, biological networks, international trade networks and others.}
Therefore, the visualization of such dynamic graphs is of great importance in revealing their temporal evolution process and many dynamic graph visualization techniques have been proposed. According to the survey by Beck et al.~\cite{beck2017taxonomy}, \textit{small multiples}, i.e., showing a sequence of static graphs, is one of the most important and basic methods for dynamic graph visualization. Prior studies~\cite{archambault2011animation,farrugia2011effective} further demonstrated that small multiples are more effective than \textit{animation}, the other basic method of visualizing dynamic graphs. 

Sometimes dynamic graphs are \textit {event-based}. In an event based dynamic graph, edges and nodes appear as individual events across time at the given temporal resolution of the data.  
An important problem is the effective selection of timeslices from the data.  In the graph drawing community, uniform timeslicing is often chosen due to its simplicity. When selecting uniform $s$ timeslices from dynamic graphs of $T$ time units, time is divided into intervals of $T/s$ and all events are projected down onto one plane for visualization. Uniform timeslicing has the advantage that each timeslice spans exactly the same interval of time.  However, it does not take into account the underlying structure of the data.

In graph mining, studies have demonstrated that the length of time intervals selected for each timeslice strongly influences the structures that can be automatically measured from the dynamic graph~\cite{krings2012effects,uddin2017optimal,karsai2014time,ribeiro2013quantifying} and affects the performance of graph mining algorithms~\cite{fish2015handling}.
Prior work in the graph mining community has explored methods for timeslicing dynamic graphs effectively.
Researchers have tried to identify appropriate window sizes for uniform timeslicing~\cite{uddin2017optimal,sulo2010meaningful} or have conducted nonuniform timeslicing~\cite{sun2007graphscope,soundarajan2016generating}.
There appears to be no single timeslicing method that is optimal for all graph mining tasks~\cite{uddin2017optimal,devineni2017one,caceres2013temporal}. Some studies have shown that different time window sizes are necessary for different analysis tasks~\cite{fish2017supervised,fish2017task} and different periods of the whole dynamic graphs~\cite{devineni2017one}.

In dynamic graph visualization, timeslice selection received little attention beyond dividing the data into uniform timeslices.
\yongwang{More specifically, how to select timeslices that are data dependent for effective visualization of dynamic graphs still remains an open problem.}
Uniform timeslicing implicitly assumes that all events will be uniformly distributed across time. However, events in a dynamic graph are rarely distributed in this way.
For example, social media streams can have a burst of edges when a topic becomes important while other time periods have very few edges.
\yongwangblue{Given the limited screen space, small multiples cannot afford a large number of timeslices and we need to carefully use the limited number of timeslices.
However,} a uniform timeslicing of such data sets will make the bursting periods suffer from visual clutter and the sparse periods remain relatively empty.

\yongwangblue{In this work, we propose a nonuniform timeslicing approach for dynamic graph visualization, which balances the visual complexity (number of edges/events per timeslice) across different intervals (Fig.~\ref{fig:teaser}). }
The timeslicing is computed based on the events present in the dynamic graph. Given a temporal resolution of the data set (e.g., second, minute, hour), we use a form of histogram equalization to make those histogram bins uniformly distributed across time.
To make viewers aware of the actual length of each interval, a horizontal bar is also shown beside the graph of each snapshot.
The major contributions of this paper can be summarized as follows:
\begin{compactitem}
	\item We propose a novel nonuniform timeslicing approach for visualizing dynamic graphs based on balanced visual complexity. 
	\item We investigate the effectiveness of the proposed nonuniform timeslicing approach through a case studies, where the common uniform timeslicing approach is also compared.
\end{compactitem}


%% file: related_work.tex
\begin{figure}[ht]
	\centering
	\includegraphics[width=0.90\linewidth]{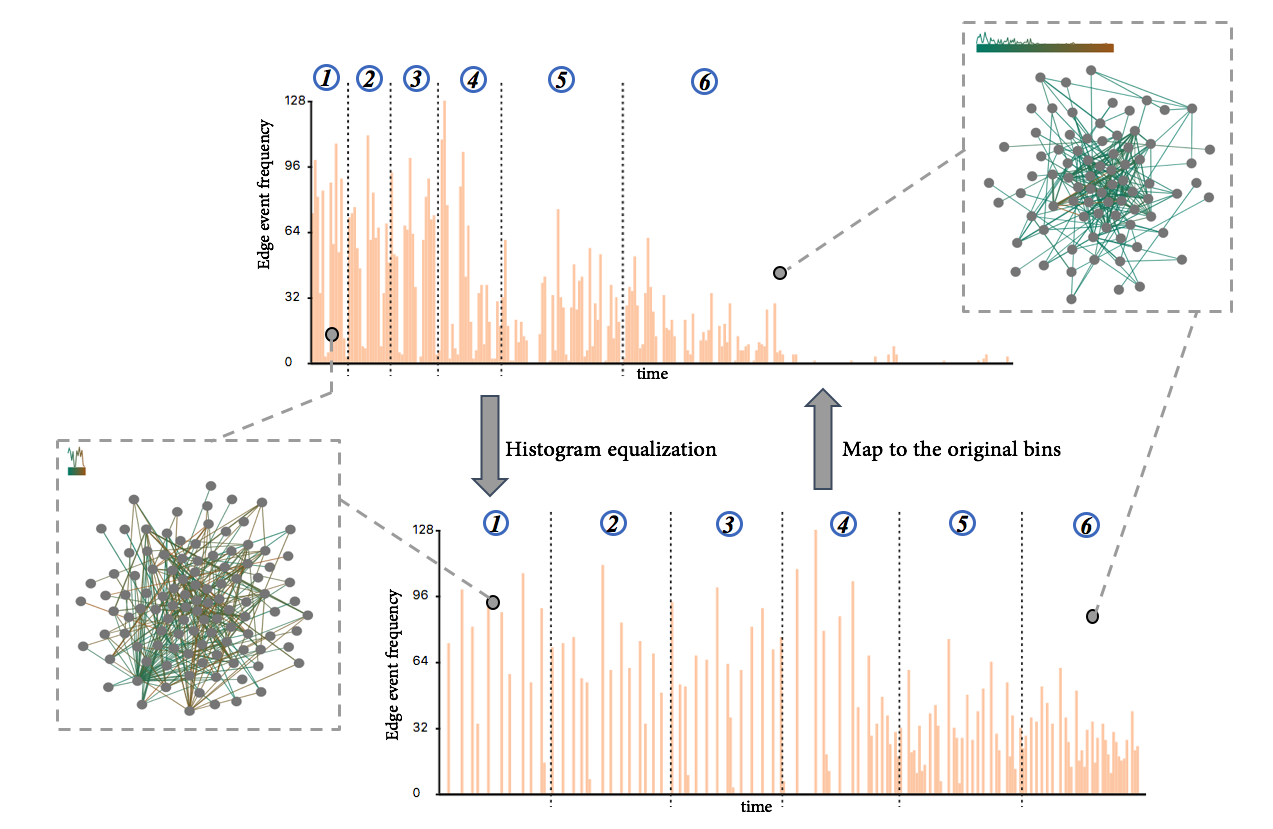}
	\vspace{-1em}
	\caption{An illustration of the proposed nonuniform timeslicing of dynamic graph based on visual complexity. The top and bottom histograms show the original histogram of \yongwang{a dynamic graph dataset} and the histogram equalization result, respectively. By dividing the bins evenly in the histogram equalization result, we balance the visual complexity across intervals, enhancing the detailed exploration for time periods with bursting edges while coarsening the periods with sparse edges, as shown by the two graphs of the first and last intervals.}
	\label{fig:teaser}
	\vspace{-1.5em}
\end{figure}

\section{Related Work}

This work is related to prior research on appropriate timeslicing of dynamic graphs and dynamic graph visualization.


{\bf Timeslicing of Dynamic Graphs:}
Prior studies in the graph mining field have been run to find ideal timeslicing methods for dynamic graphs to improve the performance of algorithms for detecting structure in dynamic graphs.
These methods can generally be classified into three categories: \textit{change point detection, minimizing the variance of a graph metric and task-oriented approaches}.
The methods based on change point detection evaluate the similarity between graphs of consecutive time units and detect \textit{change points} along time to divide the whole time range~\cite{sun2007graphscope}.
The variance-based approaches~\cite{soundarajan2016generating,uddin2017optimal,sulo2010meaningful} mainly determine the suitable timeslicing through minimizing the variance of certain graph metrics such as node degree, node positional dynamicity, etc. Other approaches~\cite{fish2017supervised,fish2017task} determine optimal timeslicing by using the accuracy of different graph mining algorithms (e.g., anomaly detection and link prediction).
%
In the visualization community, a fixed interval (e.g., one day, one month and one year) is often used to divide the graph into slices~\cite{van2016reducing,bach2014graphdiaries,shi2011dynamic,rufiange2013diffani}. 
However, we are not aware of methods that perform a nonuniform timeslicing for dynamic graph visualization based on graph structures across different intervals.

\textbf{Dynamic Graph Visualization:}
Dynamic graph visualization aims has been extensively explored in the past decades~\cite{beck2017taxonomy,17Bach,kerracher2014design,kerracher2015task,Temporal_Multivatiate_visualisation}.
Animation is the most natural way to visualize dynamic graphs, as it directly maps the evolution of the graph to an animation result~\cite{beck2017taxonomy}. 
Prior work of this type mainly attempted to preserve the mental map (i.e. the stability of the drawing) in dynamic graph visualizations~\cite{12ArchambaultGD,archambault2016can}, which are conducted through spring algorithms on the aggregated graph~\cite{huang1998line,01Diehl,02Diehl} or linking strategies across time~\cite{erten2003graphael,forrester2004graphael,baur2008dynamic,11Mader,simonetto2017drawing}.
However, animation is often less effective for long dynamic graphs~\cite{tversky2002animation}, as viewers need to memorize the dynamic evolution of a graph and check back and forth to compare different graph snapshots~\cite{bach2014graphdiaries}.
The small multiples visualization is the other major way to visualize the temporal evolution of dynamic graphs, which shows a sequence of static representations of the graph at different time intervals~\cite{kerracher2014design,kerracher2015task,beck2017taxonomy}.
Prior work has shown that the small multiples visualization is more effective than animation~\cite{archambault2011animation,farrugia2011effective} in terms of a quick exploration of the temporal evolution of dynamic graphs.
Its major limitation is the visual scalability due to the limited space~\cite{beck2017taxonomy}. 
Our approach, belonging to the small multiples visualization, assigns nonuniform time ranges for each snapshot based on the visual complexity, which partially mitigates the visual scalability issue of small multiples.

\begin{figure}[t]
\centering
  \includegraphics[width=0.95\linewidth]{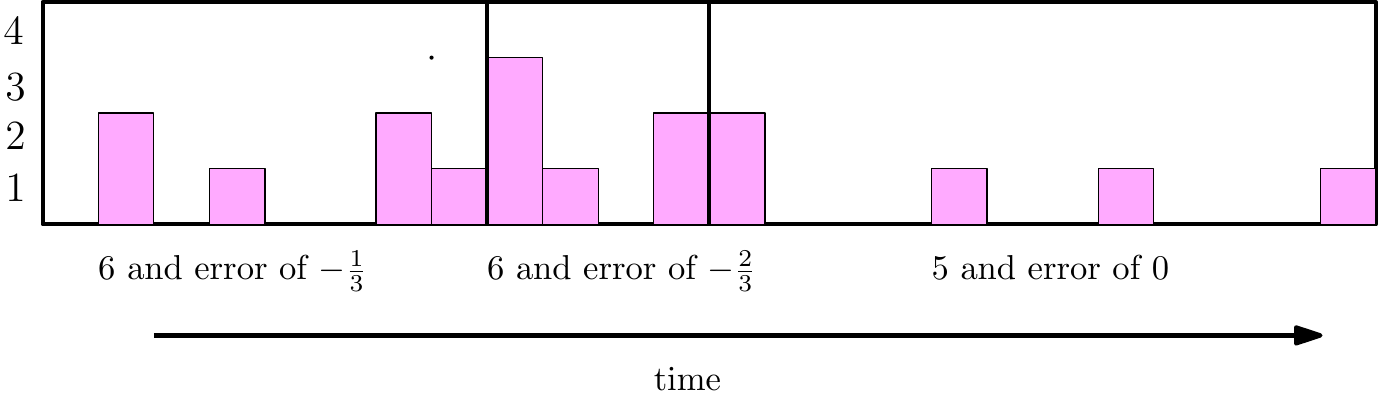}
  \vspace{-1em}
  \caption {Equal Event Partitioning.  The edges are sorted in order of earliest to latest.  Given a number of timeslices, in this case $3$, a target number of events is selected per timeslice ($\frac{17}{3}$ or $5.66$).  The algorithm counts off events, independent of timeslices, in order to fill the timeslices.  
  Equal event partitioning distributes the error evenly through time instead of giving all the error to the last timeslice.}
  \label{equEventPartFig}
  \vspace{-0.5em}
\end{figure}

\begin{figure}
\centering
   \includegraphics[width=0.95\linewidth]{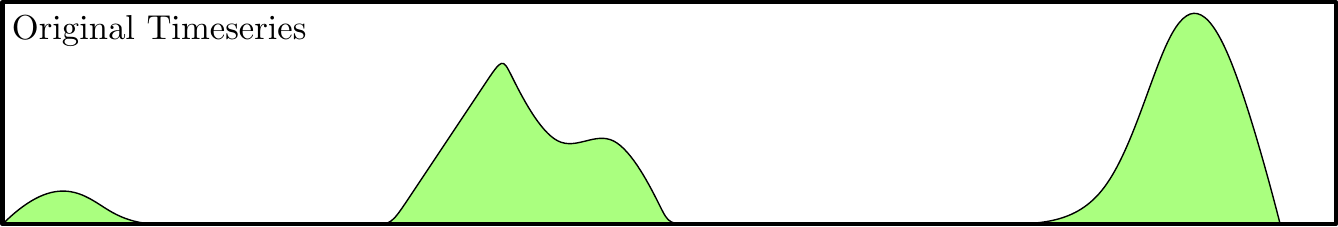}
   \includegraphics[width=0.95\linewidth]{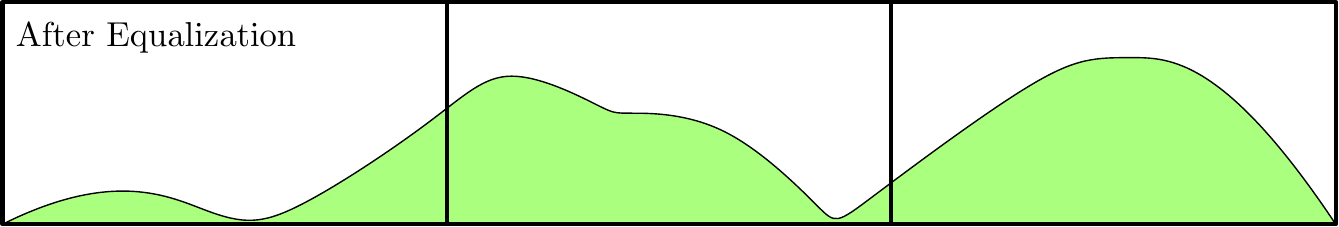}
   \includegraphics[width=0.95\linewidth]{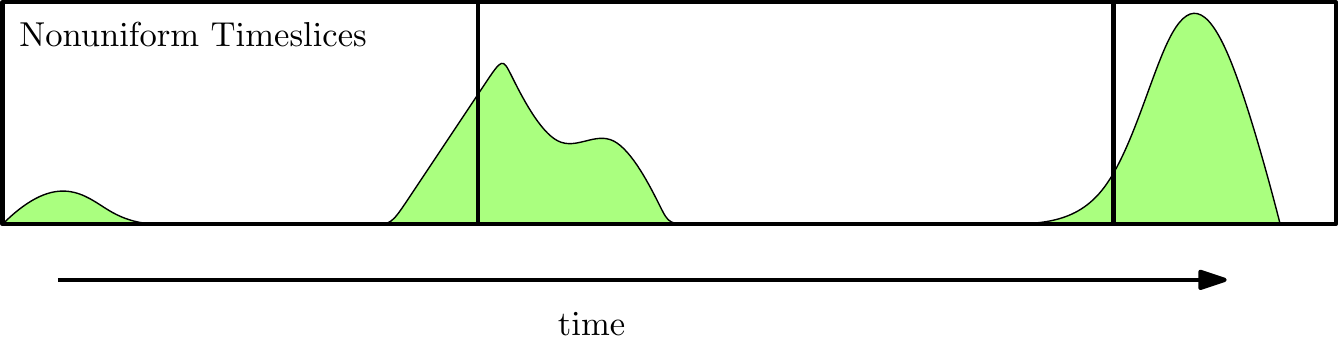}
   \vspace{-1em}
   \caption{Histogram Equalization of Events. Histogram Equalization  is adapted to event streams for dynamic graphs. The event distribution is transformed by histogram equalization  to accentuate bursts.  Regular timeslices are taken in the transformed space.  In the untransformed space, this results in a nonuniform timeslicing that accentuates bursts in the data set and skips over areas of low activity.}
   \label{histEquFig}
   \vspace{-1.2em}
\end{figure}

%% file: prob_def.tex

\section{Problem Definition and Notations}
We formally define a dynamic graph and nonuniform timeslicing of dynamic graphs according to prior work~\cite{simonetto2017drawing}.
For a dynamic graph defined on a node set $V$ and edge set $E \subseteq V \times V$, 
an edge in this dynamic graph $e_{ij}$ that appears at a time $t$ for a duration $d$ is a \textit{temporal edge}, denoted as $(e_{ij}, t, d)$, where $e_{ij} \in E$, $i,j \in V$, $t \in [0, T]$. In this paper, we choose a fixed small duration $d$ for each edge, which can also be referred as $(e_{ij}, t)$. 
Therefore, a \textit{dynamic graph} is a set of time-stamped edges that are ordered by their time stamp $t$, as defined as follows:

\begin{equation}
\langle V,E,T \rangle = \{(e_{ij},t) | E \subseteq V \times V, t \in [0,T] \}.
\end{equation}

%

Our definition does not consider timeslices as a basic unit of the dynamic graph.
The \textit{temporal resolution} of a dynamic graph is the minimum, positive distance in time between two events based on the accuracy of the time measurements. For example, edges could have an accuracy down to a day or down to a second. This temporal resolution is an important factor in our approach.  

A \textit{timesliced dynamic graph} $\Gamma = (G_1, G_2, ..., G_k)$ is a sequence of static graphs computed on $\langle V,E,T \rangle$ by dividing $[0,T]$ into intervals and projecting all temporal edges in a given interval down onto a 2D plane. Therefore, a \textit{timeslicing} $S$ on the time $[0,T]$ is: 

\begin{equation}
	S = [0, t_1), [t_1, t_2), [t_2, t_3), ..., [t_{k-1}, T],
\end{equation}
and 
\begin{equation}
	G_l = (V_l, E_l),
\end{equation}

\begin{equation}
	V_l \subseteq V,
	E_l = \{(e_{ij},t) | t_{l-1} \leq t < t_{l} \}, l = 1, 2, ..., k.
\end{equation}
where $t_0 = 0$, $t_k = T$, $E_{l}$ represents all the edge instances within the $l$-th time interval and
$k$ is the total number of time intervals we want to use for showing a dynamic graph.
If all time intervals $[t_{l-1}, t_l)$ have uniform duration, it is a \textit{uniform timeslicing}.  
Otherwise, it is a \textit {nonuniform timeslicing}. 

%% file: method.tex
\section{The Proposed Method}
In this work, we develop timeslicing methods to optimize the visualization of dynamic graphs. Specifically, we aim to have timeslices of uniform visual complexity and two methods will be introduced.

\subsection {Nonuniform Timeslicing Methods}
\label{sec-nonuniform-timeslicing}

Our definition of visual complexity in this paper is based on the number of events (in our case edges) projected into one static graph $G_i$ of the timesliced dynamic graph.
If the variance of the number of events between timeslices is small, all static representations of the graph have a similar number of events and are equally complex. 
Otherwise, a large variance will indicate that some timeslices are more visually cluttered, making it difficult to read the graph during bursts in the event stream.
Thus, the goal is to find a nonuniform partition of $[0,T]$ whereby each graph $G_i$ has approximately the same number of events. 
We accomplish this via selecting nonuniform intervals of time $[t_{l-1}, t_l)$ for which the events projected onto the graph $G_l$ are equally distributed.

The problem of computing a uniform distribution of events within each timeslice has a strong relationship with problems in image processing~\cite{histEqu}. All our methods to select a nonuniform timeslicing of a dynamic graph are inspired by image processing approaches originally designed to either enhance contrast or reduce errors in a digital image.
In essence, \emph{bursts of edges in the event stream will be given more emphasis through additional timeslices, while areas of the dynamic graph where there are few events will have few timeslices to represent them}.


In both visualization and graph mining community, a question that
is often posed is what is the optimal number of timeslices that should
be selected for a particular dataset. In a visualization context, we are
frequently limited by the screen space available. Our approach is to select
timeslices according to our definition of visual complexity given
the budget of timeslices.

To ensure that the number of timeslices does not have an effect on the layout and that our techniques are comparable, we use the DynNoSlice algorithm~\cite{simonetto2017drawing} to draw the graph once in the space time cube. All of our techniques are applied to this same drawing in 2D + time, making them comparable.

\subsubsection{Equal Event Partitioning}

The most basic way to ensure a uniform distribution of events is to place the events in order and count them until a specified number of events is reached. 
More specifically, given $|E|$ events in the dynamic graph and a budget of $k$ timeslices, we can simply create a new timeslice every $|E|/k$ events. Fig.~\ref{equEventPartFig} provides an overview of equal event partitioning.
If this method is applied directly, the error can accumulate
as fractional events cannot be assigned. Inspired by dithering~\cite{Floyd1976},
we propagate the negative or positive error closest to zero based on if we withhold from or assign an event to the next timeslice.

The strength of this method for nonuniform timeslicing is that it is very simple to implement and does ensure a uniform distribution of events across all timeslices.
But its main disadvantage is that it does not use any information about the temporal resolution of the dataset and only considers events in sequence.
It may also combine edges that are distant in time into one timeslicing. Thus, a histogram
equalization based approach is further proposed.

%
%

\subsubsection{Histogram Equalization on Events}

%

In image processing, histogram equalization can be used to enhance the contrast of images~\cite {histEqu}. Histogram equalization considers a histogram of all intensity values of a greyscale image (for example $[0,255]$) and transforms the histogram by rebinning it, so that the difference between the number of elements in each bin is reduced. 
Intuitively, the algorithm reduces or removes bins where the histogram values are low and devote more bins to areas where the histogram values are high, resulting in an image of higher contrast.

We adapt histogram equalization to process streams of edges in dynamic graphs as shown in Fig.~\ref{histEquFig}. The algorithm starts by considering a histogram where the bins are set to a temporal resolution of the dataset greater or equal to the finest temporal resolution.  The histogram represents the number of events occurring at given times across $[0,T]$.
Given this histogram with $B+1$ bins, where $E_i$ is defined to be the number of events that occur in bin $i$, we can define the probability distribution as $p(i)$, where:

\begin {equation}
  p(i) = |E_i|/|E|
\end {equation}

The cumulative distribution function $P(i)$ can then be defined as

\begin{equation}
   P(i) = \sum_{j = 0}^i p(j)
\end{equation}

We can now apply a form of histogram equalization  to transform the histogram of events into a new histogram of events $s_0, s_1, s_2, \dots s_B$:


\begin {equation}
s_i = \lfloor (T - 1) \sum_{j=0}^i p(j)\rfloor =\lfloor (T - 1) P(i)\rfloor
\end {equation}

This transformed version of $[0,T]$ accentuates bursts in the event stream and diminishes areas of low activity. If one were to watch the graph as a video, areas of bursty activity in the graph would be played in slow motion while areas of inactivity would be played in fast forward. In our approach, we uniformly sample this transformed histogram into $k$ intervals, devoting more timeslices to areas of high activity, as shown in Fig.~\ref{fig:teaser}. 

According to our experiments, if a fine-grained temporal resolution is used, the timeslicing results of histogram equalization of events and equal event partitioning are quite similar. However, if the data is recorded at coarser resolutions (e.g., month or year), histogram equalization better preserves the data granularity. Therefore, only histogram equalization of events is used in this paper.

\subsection{Visualization}

The graph drawing of dynamic graphs is not the focus of this paper, so we directly use DynNoSlice~\cite{simonetto2017drawing}, which allows us to use the same space-time cube to draw and compare the graph visualization results by uniform and nonuniform timeslicings.

As the intervals of events for nonuniform timeslicing are not of equal duration by definition, we add a small glyph, consisting of a bar and line chart, to explicitly show the time range and edge event frequency of each interval, as shown in Fig.~\ref{fig-rugby-case-study}. To further reveal the detailed time information of each edge, a color mapping from teal to brown is used to encode the time order using a color time flatting approach~\cite{17Bach} in each interval. For edges representing multiple edge events, their color is mapped to the median time of the events and the edge width indicates the number of events.

%% file: case_study.tex
\section{An Case Study on Rugby Dataset}
\label{sec-rugby-case}
We conduct a case study on Rugby Dataset~\cite{simonetto2017drawing} to demonstrate the effectiveness of our technique.
It contains over 3000 tweets between the 12 Rugby teams in the Guinness Pro12 competition.
Each tweet includes the involved teams and the accurate time stamp with a precision of one second. 
Fig.~\ref{fig-rugby-case-study} shows the timeslicing results generated by uniform timeslicing and the proposed nonuniform timeslicing approach.
Both techniques divide the whole dynamic graph into the same number of intervals (i.e., 12) for a fair comparison.


Uniform timeslicing divides the whole time range (Sept. 1st, 2014 to Oct. 23rd, 2015, 418 days in total) into 12 intervals of around 35 days each, as shown in Fig.~\ref{fig-rugby-case-study}a. 
The visual complexity across different intervals varies significantly. For example, Intervals 1, 2, 3 and 9 have sparse edges and Interval 9 even contains two disconnected graph components, revealing the infrequent interactions between the rugby teams in these time period.
But some intervals like Intervals 11 and 12 of Fig.~\ref{fig-rugby-case-study}a 
have dense interactions between the rugby teams. 
There are several bursts in Intervals 11 and 12 (indicated by the top left line charts)
and it is difficult to tell their order and structure, since uniform timeslicing does not take features of the data into consideration and often generates graphs with highly-aggregated edges for intervals of dense edges.

On the contrary, the proposed nonuniform timeslicing approach generates a sequence of graph snapshots with a more balanced visual complexity in terms of the number of edges in each interval, as shown in Fig.~\ref{fig-rugby-case-study}b. It is still easy to recognize the overall trend of interactions among rugby teams with the help of the time range bars in the top left corner. For example, the long time range bars in Intervals 1, 2, 3 and 8 of Fig.~\ref{fig-rugby-case-study}b indicates that the interactions among the rugby teams in those time periods are infrequent. More specifically, Interval 8 (late May to late August) of Fig.~\ref{fig-rugby-case-study}b has the longest time range bar, which corresponds to the summer break where there are no fixtures.
However, at the beginning of this interval, there is a burst (teal colored edges) which corresponds to the date of the Grand Final between Munster (\textit{mu}) and Glasgow (\textit{gl}) at the end of the season in 2015. The final is not easily visible in uniform timeslicing because uniform timeslicing does not accentuate it.

\yongwang{More interesting findings can be revealed by the nonuniform timeslicing approach, when there are a series of bursting edge events. For example,}
as the season begins, a number of bursts occur indicated by the short time range bars of Intervals 9-12 of Fig.~\ref{fig-rugby-case-study}b.
The nonuniform timeslicing approach is able to better accentuate certain details. For example, ``scarlets\_rugby'' (Node \textit{sc}) communicated the most with the team ``dragonsrugby'' (Node \textit{dr}) in late August, then interacted the most with the team ``glasgowwarriors'' (Node \textit{gl}) in early September, and further switched to mainly contact the team ``ulsterrugby'' (Node \textit{ul}) in late September, as demonstrated by the thickest edges linked to Node \textit{sc} in Intervals 9-11 of Fig.~\ref{fig-rugby-case-study}b.
August (Interval 9) corresponds to just before the beginning of the season. Posting activity around preseason fixtures involving Scarlets (\textit{sc}) and Dragons (\textit{dr}) as well as Edinburgh (\textit{ed}) and Ulster (\textit{ul}) are the two most prominent edges in this interval.
Scarlets-Glasgow (Interval 10) and Scarlets-Ulster (Interval 11) correspond to the first two fixtures for Scarlets in the 2015-16 season and therefore are the first two bursts of activity.
The order of these bursts is apparent because they are given separate timeslices in nonuniform timeslicing, which, however, are compacted within a single interval (Interval 11 of Fig.~\ref{fig-rugby-case-study}a) in nonuniform timeslicing.

%% file: conclusion.tex
\section{Conclusion and Future work}
In this paper, we present a nonuniform timeslicing approach for dynamic graph visualization, which can balance the visual complexity across different time intervals by assigning more intervals to the periods with bursting edges and less intervals to the periods with fewer edges.
A case study on a real dynamic graph (i.e., the Rugby Dataset)
shows that it can achieve similar visual complexity across different time intervals for a dynamic graph and better visualize \yongwangblue{the time ranges with edge bursts}.

However, several aspects of the proposed nonuniform timeslicing approach still need further work.
First, the number of intervals is empirically selected.  
Prior studies (e.g., \cite{sulo2010meaningful}) have explored empirical methods to determine the suitable number of intervals for graph mining tasks, but has not yet investigated from the perspective of graph visualization.
\yongwangblue{Also, we define the visual complexity as the number of edges/events per timeslice. Other definitions of visual complexity for dynamic graph visualization can be further explored.
Furthermore, our case study shows that our non-uniform timeslicing approach can better visualize time periods with bursting edges. However, it remains unclear which detailed graph analysis tasks can benefit from the non-uniform and uniform timeslicing approach, which is left as future research.
 }

%



